\documentclass{article}
\usepackage{geometry}
\usepackage{graphicx}
\usepackage{amsmath}
\usepackage{cite}
\usepackage[french,english]{babel}
\usepackage{amssymb}
\begin{document}

\title{Anti-photon.}

\author{Jacques Moret-Bailly.}

\maketitle

\begin{abstract}
Quantum electrodynamics corrects miscalculations of classical electrodynamics, but by introducing the pseudo-particle “photon” it is the source of errors whose practical consequences are serious. Thus W. E. Lamb disadvises the use of the word “photon” in an article whose this text takes the title. The purpose of this paper is neither a compilation, nor a critique of Lamb's paper: It adds arguments and applications to show that the use of this concept is dangerous while the semi-classical theory is always right provided that common errors are corrected: in particular, the classical field of electromagnetic energy is often, wrongly, considered as linear, so that Bohr's electron falls on the nucleus and photon counting is false. Using absolute energies and radiances avoids doing these errors. 
Quantum electrodynamics quantizes ``normal modes'' chosen arbitrarily among the infinity of sets of orthogonal modes of the electromagnetic field. Changing the choice of normal modes splits the photons which are pseudo-particles, not physical objects.
Considering the photons as small particles interacting without pilot waves with single atoms, astrophysicists use Monte-Carlo computations for the propagation of light in homogeneous media while it works only in opalescent media as clouds. Thus, for instance, two theories abort while, they are validated using coherence and Einstein theories, giving a good interpretation of the rings of supernova remnant 1987A, and the spectrum found inside. The high frequency shifts of this spectrum can only result from a parametric interaction of light with excited atomic hydrogen which is found in many regions of the universe.
\end{abstract}
\section{Introduction}
\label{intro}
As the article of W. E. Lamb \cite{WLamb} of which it uses the name, this article warns the non-specialists against the use of the word “photon”. For this purpose:

- It is initially shown that a (semi-)classical electrodynamics founded on rigorous mathematics, for example by using the mathematical, general definition of the modes of a linear field, interprets the experiments always correctly.

- While W. E. Lamb et al. are mainly interested in laboratory experiments, writing, for instance {\it The free-electron laser is an excellent example of how the \textquotedblleft photon-picture can obscure the physics \textquotedblright} \cite{WLamb2} we develop the example of astrophysics in which the erroneous use of the photon, regarded as a real particle, invalids efficient theories for the benefit of unfounded, too wonderful theories.

The usual description of classical electrodyamics fills often books, mixing strange concepts as ``radiation reaction'' with poorly defined concepts as ``optical modes''. This lack of rigour has fed critics of the semi-classical theory: Considering wrongly the field of density of electromagnetic energy as a linear field, the classical theory fails to explain experiments using photon counting, so that quantum electrodynamics seems better. On the contrary, quantum electrodynamics is unable to interpret the too fast start up of a laser.

In the whole paper, we apply strict mathematics to few, well defined concepts.

Subsection \ref{linmax} shows that the linearity of Maxwell's equations in the vacuum may be extended in matter.

Subsection \ref{modes} uses this linearity to define precisely the vector space of the solutions of Maxwell's equations, to introduce its metric from energy, obtaining an infinity of sets of orthogonal modes among which ``normal modes'' are chosen.

Subsection \ref{residuel} reminds that the small, relatively distant electric charges introduced by atomic theory are unable to absorb completely the fields emitted by other similar charges, so that we live in a residual field (that the charlatans claim to detect).

Subsection \ref{Planck1911} reminds that Planck proposed to correct his initial formula to obtain the absolute electromagnetic field from which is defined the correct field of electromagnetic energy.

Subsection \ref{Einstein} uses Einstein's theory of light-matter interactions without any other hypothesis, in particular to explain the too fast start up of the lasers.

Subsection \ref{param} studies a parametric light-matter interaction in which the coherence is kept by the use of incoherent light rather than by the short pulses used by G. L. Lamb.

\medskip
A correct use of the photon concept is so difficult that, joking, W. E. Lamb suggested ``that a licence be required for use of the word photon''. Section 3 describes errors commonly done using the photon, in particular their disastrous consequences in astrophysics.

Subsection \ref{Monte} shows how the standard use of Monte-Carlo computations applied to the photons propagating in a resonant gas, in place of Einstein theory, destroys beautiful explanations of the generation of dotted rings.

Subsection \ref{multip} introduces a multiphotonic parametric scattering which explains the observed weakness of very hot stars.

Subsection \ref{spont} shows how the rejection of parametric interactions of light with matter breaks the scale of astronomical distances and founds the big-bang theory.

\section{Recalling of semi-classical electrodynamics.}
The needed tools are:

The principles of thermodynamics.

A set of equations named ``Maxwell's equations in the vacuum'' which are linear equations describing the propagation of the electromagnetic field in an unlimited space-time, without matter. 

The equation setting the density of electromagnetic energy as a quadratic function of the field.

The equations giving the field emitted by an accelerated electric charge or a small set of charges (multipole). Einstein \cite{Einstein1917} showed that the emissions of the field result from an amplification of an initial, exciting field.

The atomic theory which sets that the electric charges (electrons, nucleus, ... ) are small at the scale of their distances.

Planck's law corrected by its author in 1911 \cite{Planck1911} and approved by Einstein and Stern \cite{Einstein1913}.

\subsection{Linearity of Maxwell's equations in matter.}\label{linmax}
The equations of the electromagnetic field in vacuum, united under the name of Maxwell's equations are linear, so that any linear combination of electromagnetic fields is an electromagnetic field. Matter is usually introduced as a continuous medium, through the approximation of permeability $\mu$ and permittivity $\epsilon$, fields often depending non-linearly on the EM field. We need to keep the linearity in matter.

We can calculate the delayed electromagnetic field radiated by an accelerated charge. By time reversal, the delayed field becomes an advanced field. By subtraction of the second system from the first, the charges are removed, while the field remains unchanged in the future. Thus, Fokker and Schwarzshild preserve the linearity of Maxwell's equations by replacing the charges by advanced fields, which modify only the boundary conditions of the equations.

\subsection{The optical modes.}\label{modes}
The solutions of a system of linear equations are represented by points of a vector space  $\mathfrak{S}$. We set that the norm of a solution of Maxwell's equations is its electromagnetic energy {\it calculated assuming no other electromagnetic field}. A mode is a set of solutions that differ only by a multiplicative real constant, it is represented by a radius of  $\mathfrak{S}$. Scalar products and orthogonality of solutions and modes are deduced from the norms.

As Maxwell's equations are defined in the whole space-time, $\mathfrak{S}$ is an infinite dimensions space. A set of orthogonal modes may be qualified ``normal''. Its choice is arbitrary \footnote{The choice of normal modes in acoustics is not arbitrary because the equations of propagation of sound are not strictly linear: the ``normal modes'' correspond to the best linear approximation of the equations.} though it may be justified in small, well defined systems: In his paper, W. E. Lamb showed on examples of laboratory experiments that finding a low dimension subspace orthogonal to the remainder of  $\mathfrak{S}$ is not easy when the use the powerful tools of SU2 algebra introduced by quantum electrodynamics seems useful.

Lasers provide almost ``monomode light rays''. They should be perfectly monochromatic beams resulting from the limitation of a plane wave by a hole and diffraction.

An astronomer is interested in the beams of light defined by the entrance pupil of a telescope and the figure of diffraction of a far point (star). These beams are not monomode because the usual light sources are time-incoherent, emitting pulses whose spectral widths corresponds to pulses long of a few meters. Such a pulse propagating in a \textquotedblleft single-mode'' beam defines an optical mode which may be qualified \textquotedblleft normal''. But, the modes observed by two close telescopes observing the same star are not orthogonal: their phases and their fluctuations (less affected by the atmosphere) are correlated. 

\subsection{The residual field}\label{residuel}
The atomic theory shows that the sizes of the electric charges (electrons, nucleus...) is much smaller than their distances. To absorb the field emitted by a charge, it is necessary to generate an opposite field using other charges. The amplitude of the electromagnetic field emitted by a charge is much larger close to it than the field emitted by other charges in similar conditions of radiation: thus, a large number of other charges is needed, it seems that it remains a ``residual field''. As the problem is very complex, the residual field cannot be computed directly.

Although we use sometimes the qualifier ``stochastic'' for the residual field, this qualifier is bad close to sources of light which increase the mean value of the residual field. It is this increase which allows some atoms to jump over a pass to a new state. 

\subsection{Planck's Formula.} \label{Planck1911} 

In response to numerous criticisms, such as non equivalence of energy of a mode to $kT_P$ for a large temperature $T_P$, Planck amended his law in 1911 \cite{Planck1911,Einstein1913}, obtaining the absolute spectral radiance inside a black body:

\begin{eqnarray}
I_\nu=\frac{h\nu^3}{c^2}\{1+\frac{2}{\exp(h\nu/kT_P)-1}\}
\end {eqnarray}

The formula remains valid if a beam escapes to a transparent medium through a small hole of the blackbody. Thus, the ``Planck's temperature'' $T_P$ of a light beam may be defined without black body, from its spectral radiance $I_\nu$ and its frequency $\nu$.

In a blackbody at the absolute zero temperature, the radiance has a mean value ${h\nu^3}/{c^2}$ which is identified with the residual radiance therefore named also ``zero point radiance''. A stranger name of this field is ``quantum field'' while it was discovered and evaluated by Planck before the birth of quantum theory.

\subsection{Basic light-matter interactions.} \label{Einstein} 
Einstein showed \cite{Einstein1917} by thermodynamics that the interaction of an homogeneous resonant medium with light may change the amplitude of the light but not the wave surfaces. This result extends to the spontaneous emission which appears as an amplification of the zero point field.

Thus, along a monomode light ray propagating in homogeneous enough media, the spectral radiance may change, increasing, in particular from ${h\nu^3}/{c^2}$, or decreasing, in particular, to this minimal value. If the radiance becomes much larger than ${h\nu^3}/{c^2}$, the light ray is called ``superradiant''. In a small volume, a ray whose radiance is maximal pumps more energy than the others, so that the others remain weak. In a large volume, only the strongest beams remain: it is the competition of the superradiant modes, observed in the lasers, where they show orthogonal modes.

\medskip
The study of the output of a starting laser tube seems to show that the amplification coefficient of the tube is twice larger for the spontaneous emission than for the established laser beam. The (semi-)classical explanation is simple, it does not require any ad-hoc concept:

When the laser works, an atom is excited by a field that it is able to absorb, that is by a ``spherical'' (dipolar, quadrupolar...) field\footnote{We use the usual word ``spherical'' although the corresponding field is not invariant by the rotations which leave a sphere invariant.}. The plane wave of the laser must be projected onto the spherical absorbable wave, and a diffracted wave. Then, the amplified wave providing an exchange of energy $h\nu$, is spherical, it must be split into the plane wave and a diffracted wave which, considering various atoms, is incoherent, thus disappears (Huygens). At the start, the atom is excited by the zero point field in the ``spherical'' mode of the atom, so that there is a single loss of amplitude in the chain of fields.

\medskip
Consider two nondegenerate levels of atoms whose populations are $N_H$ in the high state of  a transition of frequency $\nu$ and $N_B$ in the low state. Thermodynamics introduces a ``Boltzman temperature'' $T_B$ verifying the equation $N_H/N_B=exp[-h\nu/(kT_B)]$. This equation gives a negative $T_B$ if $N_H$ is larger than $N_B$, that is if the populations are inverted. In this case, the interaction of light with the atoms is an amplification. The second principle of thermodynamics applies if $T_B$ has a positive value, significant in thermodynamics. Thus light may be amplified, a laser may work without inversion of the populations.

\medskip
The density of electromagnetic energy is proportional to the square of the electromagnetic field, so that it is necessary to take into account all electromagnetic fields, in particular to use the absolute values of the fields. Too many textbooks compute the energy radiated by a dipole omitting to precise that the computation is purely mathematical, false if there are other fields.

Many authors wrote that Bohr's electron loses energy and falls to the nucleus. Taking into account the zero point field, its orbit is almost stable, fully stable if the fluctuations of the zero point field are taken into account to introduce the Lamb's shift.

A light beam received from a cold region has a spectral radiance $h\nu^3/c^2$ in the average, proportional to the square of the zero point field $Z$. A source (amplifier) of light increases the field to $Z(1+a)$, using a flux of energy proportional to $Z^2(1+a)^2-Z^2=a^2Z^2+2aZ^2$. If $a$ is much lower than 1, the first term of the result is negligible, so that the exchange of energy is proportional to the increase of the field $aZ$, not to the square $a^2Z^2$ of this increase. This computation applies to the absorption of a weakly amplified zero point field in a photon counting experiment: the signal is proportional to the decrease of the field down to $Z$, not to its square.

Many experiments study, using photon countings, the correlation of the interferences of two incoherent sources. Supposing that the signal of the photocells is proportional (as usual) to the square of $aZ$, the computation provides a wrong result in which the contrast of the fringes is too low. 

\subsection{Parametric interactions using incoherent light.}\label{param}
These interactions are described here because the preservation of the coherence does not require crystals or lasers only available in labs, so that these interactions may be frequently observed in the nature.

The theory of refraction is modified by replacing the coherent Rayleigh scattering which generates the refraction, by a scattering at a shifted (Raman) frequency.

\medskip
The atoms located between two wave surfaces, separated by $\epsilon$ small, are dressed by an electromagnetic field represented by an amplitude $A=A_0\sin(\Omega t)$. For refraction, these atoms radiate a field that, according to Huygens, reconstructs a wave surface. The out of phase component radiated by the atoms is: $A=K\epsilon A_0\sin(\omega t)$, where $K$ is a diffusion coefficient. The resulting field is
\begin{eqnarray}
A=A_0[\sin(\Omega t)+K\epsilon \cos(\Omega t)]\nonumber\\A\approx A_0 [\sin(\Omega t)\cos(K\epsilon)+\sin(K\epsilon)\cos(\Omega t)]\nonumber\\=A_0 \sin(\Omega t-K\epsilon).
\end {eqnarray}
This result defines the index of refraction $n$ by the identification:
\begin{eqnarray}
K=2\pi n/\lambda=\Omega n/c . \label{indice}
\end {eqnarray}
\medskip
Atoms can radiate a higher (anti-Stokes) frequency, whose in-phase component introduces a diffusion coefficient $K'$. With $\omega, K' > 0$:
\begin{eqnarray}
A=A_0[\sin(\Omega t)+K'\epsilon \sin((\Omega+\omega)t)]
\nonumber\\ 
A=A_0\{\sin(\Omega t)+K'\epsilon[\sin(\Omega t)\cos(\omega t)+\sin(\omega t)\cos(\Omega t)]\}
\nonumber\\
=A_0\{[1+K'\epsilon\cos(\omega t)]\sin(\Omega t)+K'\epsilon\sin(\omega t)\cos(\Omega t)\}. 
\end{eqnarray} 
As $\epsilon$ is infinitesimal and $K'$ small, $[1+K'\epsilon\cos(\omega t)]$ is close to 1. Assuming that $\omega t$ is small:
\begin{eqnarray} 
A\approx A_0[\sin(\Omega t)+\sin(K'\epsilon\omega t)\cos(\Omega t)]\nonumber\\ 
A\approx A_0[\sin(\Omega t)\cos(K'\epsilon\omega t)+ \sin(K'\epsilon\omega t)\cos(\Omega t)]\nonumber\\A=A_0\sin[(\Omega+K'\epsilon\omega)t.\label{eq5} 
\end{eqnarray} 
$K'\epsilon$ is an infinitesimal term, but the hypothesis $\omega t$ small requires that the Raman period $2\pi/\omega$ is large in comparison with the duration of the experiment $t$.

Stokes contribution, obtained replacing $K'$ by a negative $K''$, must be added. Assuming that the gas is at equilibrium at temperature $T$, $K'+K''$ is proportional to the difference of populations in Raman levels, that is to $\exp[h\omega/(2\pi kT)]-1 \propto \omega/T$.

$K'$ and $K''$ obey a relation similar to relation \ref{indice}, where Raman polarisability which replaces the index of refraction is also proportional to the pressure of the gas $P$ and does not depend much on the frequency if the atoms are far from resonances; thus, $K'$ and $K''$ are proportional to $P\Omega$, and $(K'+K'')$ to $P\Omega \omega/T$. Therefore, for a given medium, the frequency shift is: 
\begin{equation} 
\Delta\Omega=(K'+K")\epsilon\omega\propto P\epsilon\Omega\omega^2/T. 
\label{redshift} 
\end{equation} 
The relative frequency shift $\Delta\Omega/\Omega$ is nearly independent on $\Omega$. It must be integrated along a path $\ell$ of the light ray, setting d$\ell=\epsilon$.

Hypothesis $\omega t$ small requires that Raman period $2\pi/\omega$ is large in comparison with the duration of the light pulses; to avoid large perturbations by collisions, the collision-time must be larger than this duration.  This is a particular case of the condition of space coherence and constructive interference written by G. L. Lamb: \textquotedblleft The length of the pulses must be shorter than all relevant time constants'' \cite{GLamb}. 

Refraction is a parametric interaction, an interaction in which the atoms return to their original state. To obtain a balance of energy with Raman resonances, so that the interaction is parametric, at least two rays of light must be involved. The coldest rays receive energy lost by the hottest. The thermal background radiation provides cold isotropic rays. Their irradiance is large.

The path needed for a given (observable) red-shift is inversely proportional to $P\omega^2$. At a given temperature, assuming that the polarisability does not depend on the frequency, and that $P$ and $\omega$ may be chosen as large as allowed by Lamb's condition, this path is inversely proportional to the cube of the length of the pulses: an observation, easy in a laboratory with femtosecond pulses, requires astronomical paths with the nanosecond pulses of ordinary incoherent light.

The 1420 MHz spin recoupling resonance of hydrogen atoms is too high, but the frequencies 178 MHz in the 2S$_{1/2}$ state, 59 MHz in 2P$_{1/2}$ state, and 24 MHz in 2P$_{3/2}$ are very convenient: The Coherent Raman Effect on Incoherent Light (CREIL) is this parametric transfer of energy between beams propagating in {\it excited} atomic hydrogen.
 
\section{Errors induced by an usual, wrong use of the photons.}\label{erreurs}
The photons result from a quantization of normal modes of the electromagnetic field. As these modes are chosen arbitrarily, a photon is not an absolute thing. It is a pseudo-particle which cannot be exported from the system for which it was designed. But in astrophysics, it is considered as a particle, all wave optics disappears...

\subsection{Neutron-like propagation of the photons, or coherent light-matter interactions?}\label{Monte}
Using Einstein's theory, now usual in laser spectroscopy, has presented great difficulty for most physicists who claimed \textquotedblleft Townes' maser will not work''. But astrophysicists continue to follow Menzel's sentence \cite{Menzel}: ''The so-called \textquotedblleft stimulated'' emissions which have here been neglected should be included where strict accuracy is required. {\bf It is easily proved}, however, that they are unimportant in the nebulae''. For Menzel, the photon is a small particle which interacts with a single atom, losing the phase of its pilot wave.

To calculate the propagation of light in a resonant medium without coherence, the astrophysicists apply the method of Monte Carlo to photons. For each atom, the interaction is modeled by a statistical law and the conditions of interaction are drawn. This method is efficient in problems where the interaction of a particle with an atom is complex, so the phase of the pilot wave of the particle does not play a significant role. The Monte Carlo gives excellent results in the calculation of the interaction of neutrons with uranium atoms, or photons with the inhomogeneities of an opalescent medium like a cloud. But wave optics, particularly the theory of refraction shows that the phase of the these photons is only lost when density fluctuations\footnote{ or phosphorescence of complex molecules} produce Rayleigh or Raman incoherent sources of blue sky. In a dilute gas, the most frequent density fluctuations are binary collisions, whose number per unit volume is inversely proportional to the square of the density. Contrary to the opinion of Menzel, it is incoherent interactions that cannot occur in the nebulae.

One could argue that the atmosphere is a transparent medium. But the refraction of light in resonant homogeneous environments such as colored liquids or glasses does not show a significant scattering.

\medskip

Str\"omgren\cite{Stromgren} has studied a model consisting of a vast cloud of very low pressure, initially cold hydrogen in which a star is extremely hot. In the vicinity of the star, hydrogen is fully ionized into protons and electrons so it is completely transparent if rare collisions are neglected. By increasing the distance $r$ to the star, traces of atoms appear. These atoms radiate and cool the gas, accelerating exponentially the production of atoms. Str\"omgren shows the formation of a relatively thin spherical shell that absorbs star's radiation and re-emits the atomic hydrogen lines into all directions. This shell is seen as a disk, particularly bright near the limb.

\medskip
In the years immediately following the explosion of supernova 1987A, a region in the shape of an hourglass and scattering fairly high light could be observed and measured by comparing travel times of direct and indirect light (''photons echo''). Later, a system of three discrete rings (pearl necklace) appeared \cite{Sugerman}. Burrows and al. \cite{Burrows} have verified that the geometry of these rings could be interpreted by a emission of the hourglass near its limbs. Thus, they likened the hourglass to a Str\"omgren shell distorted by variations in gas density. But the rings are very thin so that, without superradiance, they have not been able to keep this interpretation.

Str\"omgren did not take into account the possibility of a strong induced emission, i.e. a strong superradiance. Let's break the Str\"omgren's shell into infinitesimal shells centred at $O$. Set $\rho$ the distance of a beam of light from the star $O$. For $\rho$ small, the angle of incidence of a beam on an infinitesimal shell is an increasing function of $\rho$, so the beam path and its amplification in each shell are increasing functions of $\rho$. If $\rho$ is larger than the radius of the outer shell, the amplification is zero. Thus there is at least a maximum amplification. Choose the smallest, setting the radius $R$ of the Str\"omgren sphere. The most intense rays, tangent to this sphere and emitted into a given direction, are generators of a cylinder of revolution seen as a circle.

As the increase of radiance of a ray is proportional to its initial radiance, the most intense rays, tangential to the sphere of radius $R$, absorb at each point more energy than others whose radiance is lower. Thus, there is a single maximum $R$. This \textquotedblleft competition of modes'' works also on the cylinder defined by a given direction, so that the circle is seen dotted. By light diffraction, the dotted circle gets the appearance of TEM(l,m) modes  of a laser for which $l$ has a nonzero value imposed for example by a circular screen.

Hydrogen emits several spectral lines, an observation in black and white blends multiple monochromatic systems. The dots (modes) are not independent because, for example, emitting a superradiant line depopulates the upper level of the transition, which favours emissions with this level as lower level. However, the complexity of the superradiant system  does not hide the analogy of the central ring of Supernova 1987A with the emission of some bad, multimode lasers.

\subsection{Parametric interactions in the Str\"omgren's shell.}\label{multip}
The absorption spectrum of atomic hydrogen obtained with a low radiance source shows only the lines of hydrogen, so that only a low fraction of the energy is absorbed.
With the high radiance rays from the star, multiphoton interactions involving a few virtual levels allow for full absorption to resonant final states. Although superradiant rays are far from achieving the radiance of the stellar emission, they depopulate the excited levels intensely, so that complete cycles of white light absorption and emission lines of hydrogen become virtual. They form a {\it multiphoton scattering induced by superradiant rays}.

Suppose that the induced scattering is efficient enough to reduce the temperature of the stellar rays to the order of magnitude of the temperature of the superradiant lines. A volume with the dimensions of the thickness of the Str\"omgren's shell has in all tangential directions a luminance a bit lower than the remaining luminance of the star. If this volume is much larger than the volume of the star, it radiates more than the star which is no longer visible.

Suppose that light crosses a huge amount of cold gas outside the shell. A parametric effect may split the frequency of an hydrogen superradiant line into the resonant frequency of some atom or molecule and the frequency of an idler. The length of the path may allow the generation of weaker, sharp lines emitted collinear with the hydrogen lines.

\medskip
Concluding, Str\"omgren's and Burrows' models explain the geometry of three rings of supernova remnant 1987A from a previously observed scattering region in the shape of an hourglass. Superradiance explains the sharpness of the rings and their discontinuities. Parametric interactions explain the high brightness of the rings by a transfer of most of the energy radiated by the star into the rings. A similar ring was observed for the planetary nebula IPHASXJ194359.5+170901 \cite{Corradi}.

Similar superradiances can probably replace gravitational effects which require proper alignment of massive stars (Einstein Cross, etc...).

\subsection{Spontaneous emission of a Str\"omgren sphere.}\label{spont}
At a distance from the star $r$ slightly less than $R$, the plasma contains few, very excited, hydrogen atoms that may slightly amplify a light ray propagating at distance $\rho$ from $O$, in particular, its Lyman alpha line. This low, \textquotedblleft spontaneous emission'' is a rapidly growing function of $r$.

At very low pressure, the incoherent scatterings are negligible. On the contrary, the parametric, coherent interactions are not disturbed by collisions, they can be intense. In particular, the scattering of light by hydrogen atoms in the 2S or 2P states exchanges energy between present radiations,  causing an increase of entropy and frequency shifts.  By this CREIL effect, the slightly amplified (\textquotedblleft spontaneously emitted'') rays absorb energy lost by the stellar radiation and lose energy absorbed by the continuum, so that their frequencies are shifted. What is the balance? The irradiance of the continuum is high, so we can assume that the balance is negative for spontaneous emission. The intensity of the amplification is an increasing function of $r$, while the redshift increases with the path to outside, therefore decreases with $r$. For $r=R$, the emitted intensity is maximum and observed at the laboratory frequency. At this point, the intensity drops sharply to 0 at higher frequencies (fig.1, a).

\begin{figure*}
\vspace*{0cm}    
\includegraphics{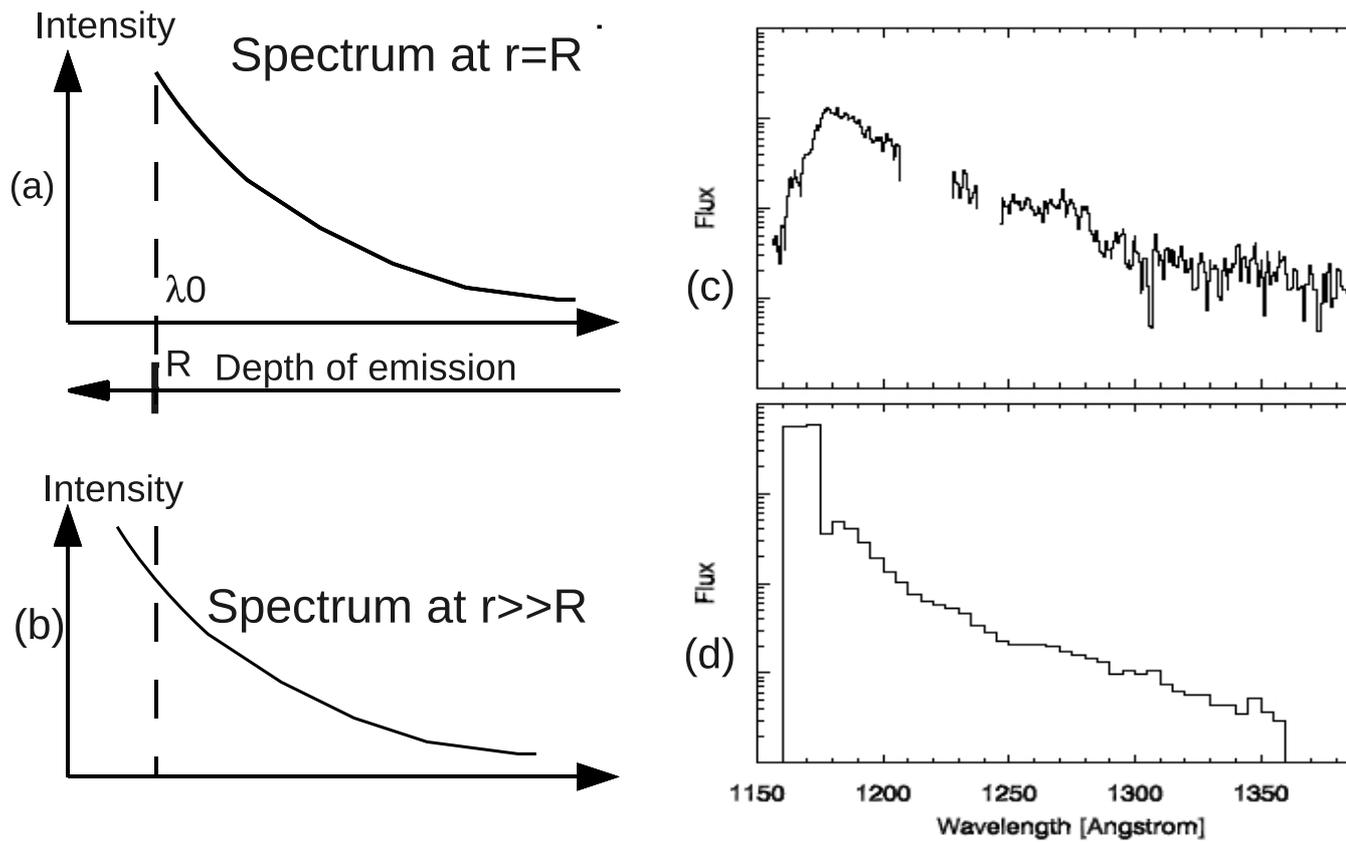}
\caption{Spectrum of weak light emitted inside the ring: (a), (b) present theory, observed at distances $R$ (a) and larger (b); (c) experimental and (d) theoretical from Michael et al. \cite{Michael}. $\lambda 0$ is Ly$_\alpha$ wavelength (122 nm).}
\label{fig:1}
\end{figure*}

 This fall, called \textquotedblleft Lyman break'' is observed in the spectra of objects called far galaxies.

For $r$ slightly above $R$, the density of excited atomic hydrogen, initially quite high to start superradiance falls fairly quickly by induced emissions. Simultaneously, the induced scattering decreases the radial propagation speed of light energy well below the speed of light. Thus, a large hot irradiance provides much energy to the spontaneously emitted rays, whose spectrum is shifted towards shorter wavelengths (fig. 1,b).

The amplifications and frequency shifts depend on $\rho$. In the observation, the spectra add for various values of $\rho$. In particular, the observed Lyman break is not very sharp. 

Michael et al. \cite{Michael} observed this Lyman alpha spectrum inside the main ring of SNR1987A (fig 1,c). They could not interpret this spectrum by an expansion of the universe applied to a very distant star observed behind SNR1987A because the solid angle of observation of the ring is relatively large. A Doppler effect of gas emitters was not plausible because the speed should be too fast. It only remained the assumption of a redshift by propagation in hydrogen, but the Monte-Carlo computation, shows a high peak, cut by the recorder (fig 1,d). The coherent, parametric, Raman effect corrects simply their spectrum improving their attribution of the large frequency shifts to an interaction with hot hydrogen, implicitely, rather than to an expansion of the universe.

\medskip
Indeed, the high redshifts interpreted by the expansion of the universe, coincide {\it always} with a presence of very hot hydrogen: 
This pseudo-expansion can be local, producing a distortion of the distance scale which inflates our charts of the galaxies and gives them their spongy aspect. The located shifts of frequency, depending slightly on the frequency, explain the spectra of stars like the quasars \cite{MBIE}, the “anomalous acceleration” of the Pioneer 10 and 11 probes, the spectra of the solar emission lines in the far ultra-violet \cite{MB0507141,MBAIP06}.

It may be a disaster for the foundations of the Big Bang theory: did a lot of people work on a theory founded on sand?

\section{Conclusion.}
Quantum electrodynamics is used to correct common errors in the use of classical electrodynamics: confusion of the field of density of electromagnetic energy with a linear field, use of a relative field to compute the density of energy, study of the absorption of a field by complex, approximate computations in place of the trivial addition of an opposite field ...

But the introduction of the the pseudo-particle “photon” is the source of errors whose {\it practical} consequences are serious.
Quantum electrodynamics is founded on quantification of normal modes whose selection among an infinity of systems of orthogonal modes is arbitrary. Thus the notion of photon should be used with so much caution that it seems desirable to reject it and, for instance, to interpret the propagation of light in a resonant medium from Einstein's theory.

\medskip
Neglecting the coherent interactions is particularly disastrous in astrophysics.

For instance, neglecting coherent interactions in the study of supernova remnant SNR1987A, wrong computations aborted precise explanations validated by coherence:

Its dotted rings result simply from a superradiant emission by the limb of the \textquotedblleft hourglass'' observed in the years which followed the explosion of the supernova.

The appearance of the rings was simultaneous with the disappearance of the star because an induced multiphotonic scattering transfers almost all radiation of the star to the rings. This explains the persistent radiation of the rings.

Michael et al. showed that the spectrum observed inside the rings must result from an interaction of light with the plasma of hydrogen, but their incoherent redshift process was too weak, so that they observed a strong peak at the resonance frequency. The coherent Raman parametric effect involving the  hyperfine frequencies of excited hydrogen atoms explains the blue-shifted peak of the spectrum, the fast decrease of its intensity at the short wavelength side, and the regular, slow decrease at the other side.

Supernova SNR1987A is an example of the utility of optical coherence in the interpretation of the aspect and the spectra of many punctuated rings, surrounding a central star possibly masked, observed in the sky. But the most important consequence is the negligence of a parametric interaction between luminous rays, catalyzed by excited atomic hydrogen; this interaction often leads to a redshift considered as due to an expansion of the universe. This pseudo-expansion may be local, 

\medskip
Coherence seems to have many other applications in astrophysics: other rings are observed, many spectra are explained.

Interpreting the strong redshifts as a consequence of the propagation of light in a medium containing excited atomic hydrogen, coherent spectroscopy undermines the foundations of the Big Bang.

\end{document}